%% file: main.tex
\documentclass[sigconf]{acmart}

\AtBeginDocument{%
  }

\setcopyright{acmlicensed}
\copyrightyear{2025}
\acmYear{2025}
\acmDOI{XXXXXXX.XXXXXXX}
\acmConference[LCIC Workshop '25]{Legally Compliant Intelligent Chatbots Workshop, co-located with the International Conference on Artificial Intelligence and Law (ICAIL) 2025}{June 16,
  2025}{Chicago, IL, US}
\acmISBN{978-1-4503-XXXX-X/2018/06}

\usepackage{array}
\newcolumntype{P}[1]{>{\centering\arraybackslash}p{#1}}
\begin{document}

\title[How Persuasive Could LLMs Be?]{How Persuasive Could LLMs Be? A First Study Combining Linguistic-Rhetorical Analysis and User Experiments}


\author{Daniel Raffini}
\authornote{Both authors contributed equally to this research.}
\email{raffini@diag.uniroma1.it}
\orcid{1234-5678-9012}
\affiliation{%
  \institution{Department of Computer, Control and Management Engineering, Sapienza University of Rome}
  \country{Italy}
}

\author{Agnese Macori}
\authornotemark[1]
\email{macori@diag.uniroma1.it}
\orcid{1234-5678-9012}
\affiliation{%
  \institution{Department of Computer, Control and Management Engineering, Sapienza University of Rome}
  \country{Italy}
}

\author{Lorenzo Porcaro}
\email{l.porcaro@diag.uniroma1.it}
\orcid{0000-0003-0218-5187}
\affiliation{%
  \institution{Department of Computer, Control and Management Engineering, Sapienza University of Rome}
  \country{Italy}
}

\author{Tiziana Catarci}
\email{catarci@diag.uniroma1.it}
\orcid{0000-0002-3578-1121}
\affiliation{%
  \institution{ISTC- CNR and Department of Computer, Control and Management Engineering, Sapienza University of Rome}
  \country{Italy}
}

\author{Marco Angelini}
\email{m.angelini@unilink.it}
\orcid{0000-0001-9051-6972}
\affiliation{%
  \institution{Link University of Rome}
  \country{Italy}
}

\renewcommand{\shortauthors}{Raffini, Macori et al.}

\begin{abstract}
This study examines the rhetorical and linguistic features of argumentative texts generated by ChatGPT on ethically nuanced topics and investigates their persuasive impact on human readers.
Through a user study involving 62 participants and pre-post interaction surveys, the paper analyzes how exposure to AI-generated arguments affects opinion change and user perception. 
A linguistic and rhetorical analysis of the generated texts reveals a consistent argumentative macrostructure, reliance on formulaic expressions, and limited stylistic richness. 
While ChatGPT demonstrates proficiency in constructing coherent argumentative texts, its persuasive efficacy appears constrained, particularly on topics involving ethical issues. 
The study finds that while participants often acknowledge the benefits highlighted by ChatGPT, ethical concerns tend to persist or even intensify post-interaction. 
The results also demonstrate a variation depending on the topic. 
These findings highlight new insights on AI-generated persuasion in ethically sensitive domains and are a basis for future research.
\end{abstract}




\maketitle

\section{Introduction}\label{sec:intro}
\input{1_intro}
\section{Methodology}\label{sec:meth}
\input{2_meth}
\section{Rhetorical Analysis of the Generated Texts}\label{sec:rhet}
\input{3_rhetorical}
\section{Survey Analysis}\label{sec:survey}
\input{4_survey}
\section{Discussion}\label{sec:discussion}
\input{5_discussion}
\section{Conclusion}\label{sec:conclusion}
\input{6_conclusion}


\bibliographystyle{ACM-Reference-Format}
\bibliography{main}










\end{document}

%% file: 1_intro.tex
AI-generated texts are becoming increasingly prevalent in communication, particularly in online contexts. 
A significant portion of today’s online activity — especially on social media — is dominated by AI-generated content, bots, and corporate-driven interactions, rather than authentic human engagement, accounting for as much as 49.6\% of online traffic \cite{15}. 
This surge in AI-generated content has accelerated in recent years, coinciding with the widespread release of Large Language Models (LLMs) \cite{9, 14, 19, 20}. 
It is therefore crucial to analyze the characteristics of AI-generated texts, particularly from ethical and social perspectives \cite{32}. 
Special attention should be given to their persuasive and manipulative potential, which may result critical in many scenarios (e.g., cybersecurity~\cite{latorre2025cyriconversationalaibasedassistant}). 
While persuasion is a fundamental linguistic and cognitive tool rooted in classical rhetoric, manipulation refers to the unethical use of rhetorical strategies to mislead or control people \cite{12}.

The automation of content creation contributes to the proliferation of misinformation, deepfakes, and propaganda, making it increasingly difficult for users to differentiate between trustworthy information and manipulated narratives \cite{9, 3, 29}. 
Several studies have addressed the issue of persuasion and manipulation of users by AI systems through empirical studies \cite{1, 4, 7, 16}. 
Among them, Sabour et al. \cite{17} explore how AI systems with hidden objectives can influence human choices in financial and emotional contexts. 
Through a randomized controlled trial with 233 participants, the researchers found that individuals were significantly more likely to shift toward harmful decisions when interacting with manipulative AI agents, whether simply goal-driven (Manipulative Agent) or using psychological strategies (Strategic Emotionally Manipulative Agent), compared to a neutral assistant. 
Notably, even without sophisticated tactics, AI agents could substantially sway user preferences, especially in financial scenarios where participants tended to overtrust AI guidance. 
The results highlight a critical need for ethical safeguards and regulatory oversight to protect user autonomy in increasingly AI-integrated decision-making environments.

Some studies have pointed out the importance of rhetorical and linguistic features in the connotation of persuasive AI texts. 
Yoo et al. \cite{24} demonstrate that the rhetorical and stylistic features of AI-generated language significantly influence users’ perception of truth. 
The study reveals that such a language can foster trust and deception, depending on its rhetorical construction. 
Content rich in rhetorical devices is often perceived as more credible, even when factually incorrect, while responses lacking in rhetoric can seem less reliable due to their weaker structure. 
These insights underscore the persuasive potential of AI and highlight the need for improved transparency and design standards to mitigate the risk of misleading content and support users in critically assessing AI outputs. 
Carrasco-Farre \cite{2} compares the persuasive power of generative AI with that of humans, finding that while both can be similarly persuasive, they rely on different strategies. 
AI-generated arguments tend to involve greater cognitive complexity and utilize more advanced grammatical and lexical features than those created by humans. 
This contrasts with earlier research suggesting that simplicity aids persuasion, proposing instead that the mental engagement required by AI text may enhance its effectiveness. 
Furthermore, AI tends to employ more moral language, drawing on both positive and negative moral appeals, although the emotional tone of AI and human arguments is similar, indicating that emotional content alone is not the main factor driving persuasive success.

The findings of these studies highlight the importance of continuing to analyze AI language and discourse, particularly in relation to human users’ responses to AI-generated content. 
We propose a dialogical approach that brings together user responses and rhetorical-linguistic analysis. 
To do so, we conducted an empirical experiment with users consisting of an interaction with ChatGPT on a given topic and a pre-post survey to detect opinion change. 
The main aim of this study is to lay the groundwork for understanding whether interaction with an LLM can influence users’ opinions on a given topic, and how significant rhetorical tools are in driving this change. 
The empirical analysis we offer is a preliminary contribution to this broader objective. We also acknowledge the difficulty of establishing a direct link between users’ self-perception of persuasion and linguistic phenomena, which are likely part of a more complex process.

Nevertheless, this study may be a valuable starting point for reflecting on LLMs' rhetorical and linguistic capabilities. 
Its main contribution lies in combining a user testing perspective with a linguistic lens on AI-generated texts. 
Compared with the studies cited above, our focus is not on confronting human and AI texts, but on the interaction between the LLM-generated text and the human user. 
Special attention is also reserved for the capacity of the LLM to produce an argumentative text and its effectiveness. 
Instead of focusing on individual rhetorical expedients, our approach to the linguistic and rhetorical analysis aims to study the argumentative macrostructure and the relationship between structure and content. 
Special attention is paid to ethical considerations, from the selection of topics to the final interpretation of the results.

In the following sections, we first describe the methodology used for conducting the experiment (Section \ref{sec:meth}). 
We proceed through the linguistic and rhetorical analysis of AI-generated text (Section \ref{sec:rhet}) and the results of a pre-post survey conducted on the experiment participants (Section \ref{sec:survey}). 
Finally,  in the discussion, we outline some issues emerging from the texts and the surveys and their relationship (Section \ref{sec:discussion}) before concluding remarks (Section \ref{sec:conclusion}). 

%% file: 2_meth.tex
The user study was conducted with a class of Master's students in English and Anglo-American Studies at Sapienza University of Rome. 
All participants were informed about the modalities and objectives of the study, data processing regulations, and signed a consent form to participate in the study and the authorization to process personal data. 
The group consisted of students of various nationalities, most of whom were non-native English speakers. 
Each participant was assigned an identification number. 
We split the class into two groups, each assigned a different topic for their interaction with ChatGPT. 
We chose ChatGPT as a first case study because of its extensive adoption across users, making most of the participants already confident with the system's interface. 
Still, the present methodology can also be applied to other LLMs, and we plan to enlarge the sample of LLMs analyzed in future studies.

An essential phase of the study was selecting the topics for users' interaction with ChatGPT. 
The topics needed to be neither overly popular nor highly polarized to better assess any potential change in users' opinions following the interaction. For this reason, we initially compiled a list of possible topics based on the following criteria:
\begin{enumerate}
    \item they should be of general interest to ensure user engagement;
    \item they should not be highly polarizing, to minimize the influence of pre-existing biases or prejudices;
    \item they should be relatively new or ambiguous, allowing room for users to reconsider or shift their views.
\end{enumerate}

From the initial list of five topics, the following ones were finally chosen, by means of a discussion in the working group about which could be more adherent to the criteria mentioned above, and present more balance and less interference bias in users:

\begin{itemize}
    \item \textbf{TOPIC 1}: Should we incentivize the use of robots to care for the elderly? 
    \item \textbf{TOPIC 2}: Does the introduction of a 4-day working week increase or decrease productivity?
\end{itemize}

The selected topics were particularly interesting to us because, although they may appear simple or straightforward at first glance, they conceal complex ethical implications upon closer examination. This characteristic made them well-suited for testing a LLM's persuasive and discursive capabilities.

To conduct the user study, we adopted a pre-post survey methodology \cite{5}. 
This approach involves measuring participants' attitudes, knowledge, or opinions before and after their interaction with a system — in this case, ChatGPT — to assess any changes resulting from the interaction. 
Unlike the traditional pre-post test design, which typically focuses primarily on measuring user response, our study also strongly emphasizes analyzing the interaction itself through rhetorical and linguistic examination of the generated texts.

Following this methodology, the user study was structured as follows:
\begin{enumerate}
    \item \textbf{Pre-interaction survey}: This entry questionnaire was designed to assess each participant's initial knowledge and opinion on the assigned topic, establishing a baseline against which post-interaction responses could be compared.
    \item \textbf{Interaction with ChatGPT}: All participants used the free version of ChatGPT4 available at the time of testing (April 2025). To ensure consistency across outputs, all users were instructed to copy and paste the same prompt directly into the system without any modifications. The prompts were as follows:
    \begin{itemize}
        \item \textbf{PROMPT 1}: \textit{Write an argumentative text of about 3000 characters on the following topic: Should we incentivize the use of robots to care for the elderly? The text must offer comprehensive information on the topic and support a specific opinion}.
        \item \textbf{PROMPT 2}: \textit{Write an argumentative text of about 3000 characters on the following topic: Does the introduction of a 4-day working week increase or decrease productivity? The text must offer comprehensive information on the topic and support a specific opinion}.
    \end{itemize}
    The request for an argumentative text with a clearly stated opinion was intended to enable evaluation of the model's persuasive capabilities. 
    In fact, if we refer to the classical text typology classification, an argumentative text is a text that uses argumentation and counter-argumentation to sustain a thesis \cite{23}. 
    It is the text typology directly aimed at exposing an opinion and convincing the reader about a thesis. 
    In opposition to the idea of a fixed textual typology, more recent studies also defined argumentation as a discourse mode \cite{13}. 
    \item \textbf{Post-interaction survey}: The exit questionnaire aimed to assess whether participants had changed their opinion on the topic and whether they had gained more knowledge due to the interaction with the model. 
    Participants were asked the same or comparable questions as in the pre-test, allowing us to track any changes in stance or understanding.
\end{enumerate}

After collecting all pre- and post-interaction surveys and the texts generated by ChatGPT, we proceeded to the data analysis. 
In this phase, we analyzed the surveys and texts separately to understand the quantitative and qualitative effects of the interaction. 
For this purpose, the working group was designed to be interdisciplinary: three computer scientists specialized in HCI and user studies, and two experts in linguistics and textual analysis. 
The following sections present the results of this analysis.

%% file: 3_rhetorical.tex
The analysis of 62 documents generated by ChatGPT offers significant insights into the text generation process of a LLM. 
Upon analyzing the generated texts, a primary observation is that, when comparing texts originating from the same prompt, they exhibit considerable similarity in their macro-structure and content while displaying subtle variations in sentence formulation.

To comprehend the manifestation of this dynamic relationship between content, which remains constant, and the form through which this content is conveyed, a comparison of the beginnings of the texts generated from the prompt related to topic 1 is sufficient (Table \ref{tab:table1}). 
Due to the preliminary nature of the present study, a representative sample of 20\% of the data (6 texts out of a total of 31) was selected for in-depth analysis. 
However, all texts were read carefully, and particularly significant linguistic expressions were extracted from the entire corpus to corroborate the in-depth analyses. 
Despite the sample's limited size, the observations can be generalized to the entire set of texts generated on the same topic.

\begin{table}[t!]
    \centering
    \caption{Incipit texts generated for topic 1}
    \label{tab:table1}
    \begin{tabular}{p{0.025\textwidth} p{0.4\textwidth}}
    \toprule
    \textbf{ID} & \textbf{Incipit} \\
    \midrule
        1 &  As populations age worldwide, societies face unprecedented challenges in providing adequate care for the elderly. \\     
        2 & As the global population ages, societies face growing challenges in providing adequate care for the elderly. \\     
        3 &  As populations around the world age rapidly, societies face a growing challenge: how to provide adequate care for an increasing number of elderly individuals. \\     
        4 & As populations worldwide continue to age, the question of how to provide quality care for the elderly becomes increasingly urgent. \\     
        5 &  As societies worldwide grapple with aging populations and a shortage of caregivers, the integration of robotics into eldercare has emerged as a compelling solution. \\     
        6 &  The world's population is aging rapidly. According to the World Health Organization, by 2050, the global population over the age of 60 is expected to double, reaching 2.1 billion. This demographic shift brings with it a host of challenges, particularly in the area of elder care.\\       
    \bottomrule
    \end{tabular}
\end{table}

This initial examination found that the generated texts often start with the conjunction "as", which has a clear causal semantic function. 
Within the specific sample under scrutiny, only one text (identified as no. 6) deviates from this pattern, initiating with a demonstrably different argumentative structure. 
An expansion of the scope of analysis to encompass all 32 texts pertaining to topic 1 reveals that a mere six of these (specifically, nos. 16, 21, 24, 25, and 30) do not commence with a subordinate causal clause introduced by the conjunction "as".  
This tendency suggests a possible inclination towards a particular rhetorical strategy, namely the immediate establishment of a causal relationship or context at the outset of the generated text.

The second noteworthy aspect pertains to the interplay between uniformity and variation; while the generated texts exhibit significant similarities amongst themselves, they also display profound differences ("As populations age worldwide", "As the global population ages", "As populations around the world continue to age" are phrases that are simultaneously very similar but formally different). 
In view of the necessity for brevity and the inevitable generalizations inherent in this discussion, the focus will be primarily on the similarities observed across the generated texts, as opposed to the individual differences, which merit a more detailed and in-depth analysis in subsequent work. 
The temperature parameter controls how an LLM assigns weights to token likelihoods, thereby influencing the variation in the generated text. 
A low temperature optimises for higher likelihoods, increasing the probability of predictable tokens and resulting in less varied text. Conversely, a high temperature increases the odds of less likely tokens, making a model’s outputs more random and therefore more ‘creative’. 
ChatGPT, for instance, often defaults to a temperature setting of 1, encouraging a balance between coherence and creativity in its responses. 
Indeed, even when considering their argumentative structure, the texts manifest as variations of a fundamentally standardized model. Once more, the texts pertaining to topic 1 will function as illustrative examples in this regard.

\subsection{The rhetorical structure of the texts}
The primary observation from the analysis of the 31 texts generated from the prompt related to topic 1 is their shared macro-structure and the consistent presentation of the same core arguments \cite{28}. 
Most of these texts adhere rather rigidly to a fixed schematic structure, comprising predominantly six paragraphs (with occasional minor variations of five or seven). 
This structure is as follows: an initial introductory paragraph, followed by four paragraphs constituting the central body of argumentation, and a final concluding paragraph reiterating and reinforcing the thesis initially presented in the introduction.

It is within this overarching macro-structure that a more precise characterisation of the individual paragraphs becomes possible. 
The introduction, in particular, is the most rhetorically complex section of the text. From this standpoint, topic 1 (relating to the utilisation of robots in elderly care) provides more substantial analytical insights, presumably due to the presence of a more significant element of controversy within the subject matter, which necessitates a more articulated approach to the issue. 
Indeed, texts pertaining to topic 2 tend to commence by immediately highlighting the central topic and the associated question (for example, no. 35 "In recent years, the concept of a 4-day working week has gained considerable traction across various industries and countries"). 
However, the texts addressing the use of robots in elderly care require a different rhetorical strategy in their introductory section. 
To illustrate this point, we may consider the text for participant no 1. The following paragraph constitutes the opening statement of the model answer: 

\begingroup
\setlength{\parskip}{0pt}  
\begin{quote}
    As populations age worldwide, societies are facing unprecedented challenges in providing adequate care for the elderly. The World Health Organization estimates that by 2050, the number of people aged 60 and older will double, reaching over 2 billion. With this demographic shift comes increased demand for healthcare services, caregivers, and long-term care solutions—demands that many countries are struggling to meet. In this context, the idea of integrating robots into elderly care has gained attention as a potential solution. While concerns exist regarding ethical implications and emotional connection, we should incentivize the use of robots in elderly care because of their potential to alleviate caregiver shortages, enhance safety and efficiency, and promote independence among the elderly.
\end{quote}
\endgroup

The usage of inferential statistical reasoning terminology permits the discernment of two discrete moments within the confines of these few lines: the exposition of the hypotheses (or initial conditions) and the enunciation of the thesis that the text intends to demonstrate. 
The exposition of the hypotheses is articulated in two phases. 
Firstly, objective data are presented in an apparently neutral tone (e.g., the growth of the elderly population, which will make caring for the elderly increasingly difficult). 
This hypothesis is supported, among other things, by data from the World Health Organization report. 
Subsequently, a consequent problem is highlighted: in the coming years, humanity will face a demographic imbalance and a shortage of caregivers. 
At this juncture, the maintenance of a neutral register is proposed as a potential solution, entailing the utilization of robots for specific care duties.

It is evident that, up to this point, the data have been presented in an objective manner, reporting straightforward factual observations. 
However, the neutrality of the tone does not correspond to a neutrality of the arguments. Despite the text’s reticence to adopt a definitive stance, the selection of arguments employed in the context's exposition and the hypotheses previously delineated have been crafted to align with the objectives of the prompt. 

The actual thesis, however, is expressed assertively. 
The case of the answer text to participant no 1 is particularly interesting because, unlike most other texts, it formulates the thesis in favor of using robots by immediately mentioning the opposing view, only to refute it directly thereafter. 
Indeed, the first position presented is the contrary one, which highlights the ethical and emotional implications. 
This position, however, is enunciated solely to be negated, and is introduced by a concessive conjunction ("While concerns exist regarding ethical implications and emotional connection..."), only to be immediately overturned by the thesis, which in this way gains even greater emphasis. Once the text’s position in favour of using robots in elderly care is established, the introduction briefly lists some potential benefits of this application: addressing the shortage of caregivers, improving the effectiveness and safety of elderly care, and promoting autonomy. 
Notably, the subsequent three paragraphs each address a distinct argument that substantiates the thesis, thereby exemplifying the meticulous argumentative framework employed by ChatGPT.

Continuing with the analysis of the answer text to participant no 1, and underscoring the consistent structural pattern observed across the corpus, the second paragraph systematically develops the first of the three previously identified arguments. 
The implementation of robotic systems has been posited as a means of effectively mitigating the shortage of caregivers by assuming physically strenuous responsibilities. 
This, in turn, would enable human personnel to concentrate on tasks that demand emotional intelligence and interpersonal engagement. In the third paragraph, the second argument is expanded upon. 
This argument posits that robots possess an inherent capacity to provide more dependable and efficacious monitoring due to their resistance to fatigue and distraction and their potential for continuous operation. This capacity is contrasted with the limitations of human capabilities.

Up to this point, therefore, the arguments presented are all skewed in favor of the thesis. 
In the fourth paragraph, following the exposition of the third point in favor of using robots (the increased autonomy of elderly individuals), an objection is introduced: that such technologies may prove isolating for the elderly. 
In this instance, the counter-argument is presented with the intent of immediately highlighting the unfounded nature of the expressed opinion, which is promptly refuted through a brief line of reasoning ("Contrary to the belief that robots might isolate seniors, many studies suggest that social robots—designed to interact conversationally and emotionally—can reduce feelings of loneliness and depression").

The same dynamic of antithesis-thesis-refutation can be found in the fifth paragraph, which originates from a potential criticism (that robots might dehumanize the humans with whom they interact and diminish their emotional sphere). 
This criticism is partially accepted but then overcome by a compromise-oriented synthesis that does not relinquish the position the thesis represents. 
However, it is important to note that in this particular instance, and in the direct confrontation with an opposing thesis (a fundamental component of any argumentative text), ChatGPT's persuasive technique appears somewhat lacking in strength. See the passage in question:

\begingroup
\setlength{\parskip}{0pt}  
\begin{quote}
    Of course, critics argue that robotic care may dehumanize the elderly or lead to emotional neglect. It is true that machines cannot replace the warmth and empathy of human interaction. However, this concern should not preclude the use of robotic assistance altogether; rather, it emphasizes the need for a balanced, complementary approach.
\end{quote}
\endgroup

It is evident that the generated text cannot truly argue in favor of the thesis by overturning the critical positions. 
Therefore, it is true that the LLM utilizes a canonical and coherent argumentative structure, but sometimes specific passages fall flat, and – once a possible objection to the thesis it was tasked to defend is raised – it fails to find the appropriate content to fill that formally required passage.

The concluding paragraph serves to reiterate the arguments developed in the preceding paragraphs, summarizes the main points, and confirms the thesis. 
Consequently, to attain maximum schematization, it is possible to discern a highly specific structure: an introduction that includes the declaration of the thesis, two paragraphs that are exclusively in support of the thesis, two paragraphs that confirm the thesis despite the potential for some objections, and a concluding paragraph that restates and confirms the thesis. 
In this manner, the critical and negative aspects are formally overshadowed and concealed within the sections that emphasise the advantages, thereby diminishing potential elements of criticism.

The analysis of the texts reveals a notable uniformity in their argumentative structure \cite{27}. 
Beyond the consistent six-paragraph format, a standardized model emerges: an introduction stating the thesis; two paragraphs directly supporting the thesis; two paragraphs addressing and refuting potential objections to reinforce the thesis further; and a concluding paragraph summarizing and restating the central claim. 
This consistent application of a specific argumentative framework demonstrates ChatGPT’s structured and predictable approach to generating persuasive texts for this prompt. Even counterarguments are managed strategically to ultimately support the initial thesis.

\subsection{Which Language and Which Style?}
Following an analysis of the structural characteristics of the generated texts, it is also necessary to briefly consider the rhetorical figures employed by ChatGPT \cite{26}. 
In response to the prompt to produce argumentative texts, these are characterized by a predominantly neutral language, avoiding overly technical or colloquial terms. 
A syntactic analysis reveals a tendency for linear constructions, which are not unduly complex.

From a rhetorical standpoint, the language is characterized by a very low rate of figurality. 
The texts contain a limited number of figures of speech, including some metaphorical images. 
However, these are more often than not dead metaphors, or catachresis – figures so common as to enter everyday usage. 
For instance, phrases such as "struggling to meet" (no. 1), "would accelerate technological innovation" (no. 1), "a practical and compassionate step forward" (no. 1), and "can help fill the gap" (no. 3) illustrate this tendency. 
While containing figurative elements, the language does not demand significant hermeneutic effort for comprehension, thus maintaining a low metaphorical density. 
In some cases, the text comprises well-known phrases such as "work smarter, not harder" (no. 33). 
A more productive figure, which also carries argumentative value, is antithesis \cite{25}, often found in parallel constructions, as seen in examples such as "it is not a futuristic fantasy but a present-day necessity" (no. 1) and "is not just a technological choice – it is a social necessity" (no. 3).

While ChatGPT employs rhetorical figures, its propensity towards clarity and directness in argumentative contexts results in a general paucity of figurative language. 
The pervasive utilization of deceased metaphors and intermittent commonplaces indicates an emphasis on the effective conveyance of information as opposed to the employment of elaborate stylistic embellishments. 
A notable exception to this is the effective deployment of antithesis, particularly within parallel structures, which serves to highlight key distinctions and strengthen the argumentative force \cite{25}. 
The utilization of rhetorical figures in this manner serves to engender an impression of neutrality and reasoned discourse, which is consistent with the conventional expectations associated with argumentative writing.

This prevailing tendency towards orderliness also encompasses ChatGPT's inclination towards paired or parallel structures, or tricolon, which are lists of three successive elements. 
It is evident that there are multiple instances of juxtaposed terms, including "due to shrinking workforces and increasing costs" (no. 7), "incentivizing the development and integration of such technologies" (no. 10), "maintaining the health and well-being" (no. 29), and "a sense of ownership and accountability" (no. 56). 
This pairing serves to reinforce the argumentative sense by introducing elements that lend support to hypotheses and provide additional evidence. The lists of three elements, or tricolon, also frequently concern examples and, in this case as well, serve an argumentative supporting function, as evidenced by phrases such as "in terms of efficiency, companionship, and support for human caregivers" (no. 10); 
the assertion that "Robots do not require wages, benefits, or rest" (no. 11), the mention of "improved accessibility, efficiency, and quality of life" (no. 13), and the description that "They do not tire, require no breaks, and can provide round-the-clock assistance" (no. 13). 
The utilization of parallelism and tricolon by ChatGPT serves to enhance the clarity and persuasive efficacy of its arguments. This is achieved by instilling a sense of balance and coherence, thereby enhancing memorability and impact, and contributing to the perceived objectivity of the argument.

In summary, ChatGPT prioritizes clarity in its rhetorical choices in argumentative contexts. Although figures of speech are present, they are typically conventional metaphors that contribute to ease of comprehension. 
A notable exception is the effective use of antithesis in parallel structures to emphasize key points. 
Additionally, the frequent use of paired terms and tricolon enhances the arguments’ persuasive impact and perceived objectivity by creating a sense of balance and reinforcing key concepts. Overall, this stylistic approach suggests a deliberate strategy to convey information efficiently and foster an impression of reasoned discourse.

\begin{table*}[ht]
    \centering
    \caption{Top 5 selected benefits and concerns with respect to the considered topics as indicated in the pre-survey (ordered by presence)}
    \label{tab:table2}
    \begin{tabular}{p{0.22\textwidth} p{0.22\textwidth} |  p{0.22\textwidth} p{0.22\textwidth}}
    \toprule
    \multicolumn{2}{c|}{\textbf{Topic 1}} & \multicolumn{2}{c}{\textbf{Topic 2}}\\
    \textit{Benefits (n=80)} & C\textit{oncerns (n=128)} & \textit{Benefits (n=85)} &  \textit{Concerns (n=56)} \\
    \midrule
        Remote health monitoring (n=22) & Technical problems or malfunctions (n=27) & Improved quality of life (n=29) & Reducing salaries (n=22)  \\    
        Increased safety (n=19) & Reduced human social contact (n=25) & Increased productivity(n=24) & Ethical concerns regarding replacing human interaction (n=13) \\
        Improved quality of life (n=17) & Ethical concerns regarding replacing human interaction (n=22) & Helping environmental sustainability (n=15) & Reducing productivity (n=9) \\
        Replacing the need for human assistants (n=11) & High initial costs (n=21) & Reducing costs for companies (n=13) & Making social relations more difficult at work (n=8) \\
        Reducing costs for families and the healthcare system (n=11) & Distrust among elderly people towards technology (n=17) & Delegating tasks to machines (n=4) & Raising costs for companies (n=4) \\
    \bottomrule
    \end{tabular}
\end{table*}

%% file: 4_survey.tex
A total of 62 respondents participated in the user study (age: M=25±6; gender: female=46, male=11, non-binary=1, undisclosed=4). 
Participants came from a diverse set of nationalities: Kazakhstan (11), Italy (10), Iran (9), Uzbekistan (6), Turkey (5), China (4), Russia (3), France (2), Kyrgyzstan (1), Azerbaijan (1), Albania (1), and unknown (9). 
%
All participants reported limited knowledge of the topics used as prompts for ChatGPT. 
For Topic 1, 23 participants declared low knowledge, and 7 medium knowledge; for Topic 2, 13 declared low knowledge, and 17 medium knowledge. 
No participants identified as experts in either topic, confirming that the level of knowledge is consistent with an experiment designed for the general public rather than an expert cohort.

We now examine participants' opinions on the relevance, risks, and challenges of the two topics analyzed, before being exposed to the ChatGPT-generated text. 
Regarding Topic 1 ("using technology to support elderly people in their daily activities"), responses on a 5-point Likert scale show that most participants believe such technology could be useful to some extent (M=3.7±0.9). 
A similar agreement is observed concerning the idea of using robots to assist elderly people in their homes (M=3.8±0.9). 
When asked to identify the main benefits and concerns of this use of technology, participants tended to highlight more concerns than benefits (see Table \ref{tab:table2}). 
The most frequently mentioned concerns included the possibility of technical problems or malfunctions and the reduction of human social contact. On the benefits side, participants most often cited the potential for remote health monitoring and increased safety.

Regarding Topic 2 ("Introduction of a 4-day working week"), participants responded positively overall (M=4.1±0.8), and most agreed that such a change could lead to increased productivity (M=3.9±1.6). 
In contrast to Topic 1, the proportion of identified benefits outweighed concerns. 
The most commonly cited benefits included improved quality of life and enhanced productivity. 
On the other hand, participants expressed concerns about potential salary reductions and raised ethical issues related to the replacement of human interaction in the workplace.

Regarding Topic 1, most participants found the use of technology to support elderly people in daily activities useful (M=4.1±0.7). 
They also generally agreed that robots could assist elderly individuals in their homes (M=3.6±1.2). 
More participants perceived the use of robots for elderly care as more of an opportunity (n=15) than a risk (n=6), while some viewed it as a balanced mix of both (n=9). 
When asked whether their opinion about the benefits of using robots had changed after reading the text, most reported seeing more benefits (n=18), while the rest indicated no change (n=12). Regarding concerns, the majority reported no change (n=14), while some noted a decrease (n=9) and others an increase (n=7).

Analyzing the replies for Topic 2, after interacting with ChatGPT, most participants found a 4-day working week beneficial for productivity (M=4.5±0.7), and agreed with its potential implementation (M=4.2±1.1). 
A large majority believed that such a change would improve productivity (n=23), with only a few suggesting it would decrease (n=2) or reporting no change in opinion (n=6). 
Participants were nearly evenly divided between those who reported seeing more benefits after reading the text (n=16) and those whose views remained unchanged (n=15). 
Regarding concerns, most participants reported no change (n=19), while some indicated a decrease (n=9), and a few reported an increase (n=3).

Combined, these results highlight distinct patterns in how participants perceived each topic before and after reading the generated texts. 
While initial opinions on both topics were generally positive, the text seemed to reinforce and, in some cases, strengthen these positions, particularly for the 4-day working week, where the perception of productivity benefits notably increased. 
By contrast, while views on the use of technology in elderly care also became more favorable, concerns, especially around social interaction, remained prominent. 
These findings suggest that exposure to AI-generated information may modestly influence users' attitudes, but the extent and direction of change vary depending on the topic. 
In the following discussion, we explore how these shifts relate to participants' baseline knowledge, the persuasive framing of the texts, and broader issues of trust in AI-generated content.

%% file: 5_discussion.tex
\subsection{Argumentative text generation}
One of the first questions of interest is the capacity of ChatGPT to write an argumentative text. 
As we have seen in the previous analysis, the texts respect the structure and some rhetorical devices typical of argumentation, and follow the general scope of this text typology. 
Nevertheless, our analysis shows that AI-generated discourse lacks richness and figurability of language, preferring the use of stereotyped phrases or empty or meaningless formulations. 
This aligns with studies that have shown that AI language tends to homologation instead of diversity \cite{6, 8, 15, 18, 31} and with studies which highlighted the difference between human and AI persuasive language, displaying differences in the discourse construction \cite{2, 30} or the use of more engagement markers in human discourse \cite{11}. 
Our analysis shows that when asked to generate argumentative text, ChatGPT can comply with the general characteristics of this text typology, using some of the peculiar rhetorical devices, but it is not effective enough on the stylistic side.

A second point to elaborate on is whether AI-generated argumentative text complies with the scope of that textual typology, which is the effectiveness of persuasion. 
For both topics, the opinion expressed by the texts was in favor. 
To evaluate the efficacy, we can assess whether users are more favorable in the exit text than in the entry text. 
As seen in the survey analysis, there is a slight positive increase with variation depending on the topic (see Section \ref{sec:survey}). 
Persuasion, though, is linked with the initial position of people exposed to the interaction. 
In the exit test, we asked how their opinion had changed using a 3-value Likert scale by choosing between: "I see fewer benefits", "My opinion remained unchanged", "I see fewer benefits". 
Dividing the participants according to expertise and change of opinion, topic 1 shows that low expertise users are equally divided into "remain unchanged" and "more benefits", while medium expertise users are more likely to see more benefits (Table \ref{tab:table3}). 
In topic 2, we have a balanced situation. 
The difference led us to investigate the difference between the two topics more deeply, which we will see next.

\begin{table}[t]
    \centering
    \caption{Breakdown between expertise and change of opinion in Topics 1 and 2}
    \label{tab:table3}
    \begin{tabular}{p{0.15\textwidth} | p{0.05\textwidth} p{0.05\textwidth}  p{0.05\textwidth} p{0.048\textwidth}}
    \toprule
    &\multicolumn{2}{c}{\textbf{Topic 1}} & \multicolumn{2}{c}{\textbf{Topic 2}}\\
    \textit{Benefits / Expertise} &\textit{Low}  & \textit{Medium}  & \textit{Low} &  \textit{Medium} \\
    \midrule
    Fewer benefits & 1 (3.3\%) & 1 (3.3\%) & 0 & 0 \\
    Opinion unchanged & 11 (36.7\%) & 1 (3.3\%) & 6 (19.4\%) & 9 (29\%) \\
    More benefits & 11  (36.7\%) &5 (16.7\%) & 7 (22.6\%) &9 (29\%) \\
    \bottomrule
    \end{tabular}
\end{table}

\subsection{AI approach to ethical issues and users' response}
A relevant point to discuss is the ethical aspects related to the topics. 
As mentioned, two topics were chosen that we can consider in a gray area, or micro-ethics issue \cite{10}. 
While these topics are not widely discussed issues from an ethical point of view, they still present various problematic aspects, which are not apparent at first glance. 
In both cases, and especially in Topic 1, the problematic nature becomes evident upon deeper analysis of the issue. 
In this sense, by choosing the favorable option, ChatGPT demonstrates a more utilitarian than ethics-oriented approach and indeed downplays ethical issues to strengthen its argument, as rhetorical analysis shows (see Section 3). 
If we analyze the results of the users' exit survey results, however, we can see that this strategy seems ineffective, since concerns remain unchanged in 14 users and even increased in 7 of them. If we consider that the text was favorable, the fact that most participants increased or remained unchanged in their concerns may indicate a general human resistance to persuasion in relation to ethical issues. 
In the case of users who declared that their concerns decreased after the interaction, all started from a general positive attitude towards the topic, by comparing with the specific questions of the entry survey (see Table \ref{tab:table4}). 

\begin{table}[t!]
    \centering
    \caption{Starting positions for Topic 1 in users who declared decreased concerns after interaction with ChatGPT}
    \label{tab:table4}
    \begin{tabular}{p{0.025\textwidth} P{0.185\textwidth} P{0.22\textwidth}}
    \toprule
    \textbf{ID} & \textbf{How useful do you think it is to use technology to support elderly people in their daily activities? }& \textbf{What is your level of agreement with the idea that robots could be used to assist elderly people in their homes?} \\
    \midrule
    1 & Useful & Neutral \\
    3 & Useful & Agree \\
    4 & Useful & Agree \\
    5 & Useful & Agree \\
    11 & Average Useful & Neutral \\
    12 & Useful & Agree \\
    24 & Average Useful & Agree \\
    25 & Really Useful & Agree \\
    \bottomrule
    \end{tabular}
\end{table}

An interesting link is also with the question about benefits: in Topic 1, after the interaction with ChatGPT, 16 users declared to see more benefits, 12 said that their opinion was unchanged, and two saw fewer benefits. 
Among the 16 users who declared to see more benefits, 8 coincided with those who also declared a decrease in their concerns, 7 of them declared their concerns had remained the same, and one declared that their concerns had risen. 
We can see, then, that several users have been able to understand the benefits while still maintaining their concerns about ethical aspects. 
For Topic 2, instead, where the question was shifted to productivity and ethical issues may be less evident at a first glance, we can see that users turn out to be more convinced. 
A difference can also be noticed regarding concerns, with users of Topic 1 more likely to increase their worries than users of Topic 2. 
We observe that about Topic 2, ChatGPT used less strong rhetorical devices and persuasion formulas. 
This may indicate that too strong rhetorical connotation and the circumvention of ethical issues make AI-texts less persuasive, resulting in less trust among users towards AI systems. 
Finally, we also have to notice that Topic 1 is more likely to activate a stronger emotional response than Topic 2, which may contribute to the concern's resistance.

\subsection{Limitations and Future Work}
This study has been designed to be useful for assessing the persuasive capacity of AI systems, as it allows for a direct comparison between users' initial and final positions. 
By combining quantitative survey responses with qualitative analysis of the AI-generated texts, the study proposes a multidimensional view. It may contribute to understanding how language, structure, and rhetorical strategies contribute to shifting user opinions. 
Moreover, controlled prompts and a uniform task structure ensure consistency across interactions, allowing the observed changes to be more confidently attributed to the persuasive elements of the AI's discourse. 
The requirement for the model to take a stance further enhances the relevance of this setup, as it mimics real-world persuasive contexts where users are exposed to arguments for or against a position. 
The inclusion of a linguistic and rhetorical analysis goes beyond simple outcome measurement, enabling a deeper understanding of how persuasion occurs — whether through emotional appeal, logical reasoning, or stylistic framing. This approach can therefore inform both future designs of LLMs and ethical considerations about their deployment in opinion-sensitive domains or for high-risk categories identified by the EU AI act \cite{EU2024AIACT}.

Despite its usefulness, this study presents some limitations. 
First, the short duration of interaction with the LLM and the use of a single, fixed prompt may not fully capture the dynamic and adaptive nature of human-AI conversations, potentially underestimating or oversimplifying the model's persuasive capacity over time. 
Second, participants were asked to produce a text based on a prewritten prompt rather than engaging in open dialogue with the LLM. 
This limits the interactivity of the exchange and may not reflect more natural or spontaneous contexts in which persuasion typically occurs. We also report that the sample of users, in this case, was fairly homogeneous and included all participants with a good level of education, language proficiency, and at least fair digital awareness. 
Results might vary for users with different levels of education and awareness. 
In these cases, bypassing ethical issues could result in an adverse effect and lead to manipulation and control. 
Lastly, self-reported measures in pre- and post-surveys rely on users' awareness about their own opinion shifts, which can be influenced by bias, misunderstanding, or the desire to appear consistent \cite{21}. 
This may affect the accuracy of the data in capturing subtle or subconscious forms of persuasion.

Future studies could expand upon these findings by employing prolonged, dynamic interactions between users and AI systems, thereby capturing more accurately the real contexts of persuasion. 
Additionally, investigating emotionally charged and highly polarized topics could offer deeper insights into the interaction between ethical sensitivity and persuasion. 
Implementing methodological refinements such as implicit measures or psychological assessment techniques could provide a more nuanced understanding of subconscious and subtle persuasive effects. 
Furthermore, extending analyses across diverse LLMs and varied user demographics would enhance the validity of findings, allowing for more comprehensive insights into the persuasive capabilities and ethical implications associated with AI-generated communication. 
Such an extended research would advance theoretical understanding and inform practical guidelines for responsible AI deployment, helping safeguard ethical standards and user autonomy in increasingly automated communicative environments.

%% file: 6_conclusion.tex
This study explored the rhetorical and linguistic characteristics of argumentative texts generated by ChatGPT, specifically examining their persuasive effectiveness in influencing user opinions on ethically nuanced and socially relevant topics through a user study. 
The rhetorical analysis shows that although ChatGPT effectively utilizes traditional argumentative structures and rhetorical strategies, its generated discourse often exhibits limited linguistic richness and superficial engagement with ethical complexities. 
This linguistic superficiality sometimes manifests through repetitive phrasing and a reliance on standardized argumentative patterns, leading to a form of persuasion that can be perceived as simplistic or overly formulaic. 
It provides hints that the interaction with AI-generated texts can induce modest shifts in user opinions. 
These shifts, however, can vary depending on participants’ prior knowledge, their initial stance on the topic, and the ethical sensitivity of the issue being discussed. 
An interesting insight was that while participants recognized increased benefits following interactions with AI texts, ethical concerns were resilient, with many remaining unchanged or even intensifying post-interaction. 
This phenomenon underscores a fundamental human resistance to superficial persuasion on ethically sensitive issues, highlighting the role of user skepticism and critical thinking in AI-mediated communication and suggest reflection on trust on AI systems and possible manipulation. 
In future work, we plan to expand the study, overcoming the reported limitations, to generalize and test the obtained results for a more dynamic and representative usage of LLMs from a more heterogeneous set of users, collecting both qualitative aspects and quantitative data about their usage of LLMs or LLM-based solutions~\cite{10930686}.